\documentclass[twocolumn]{aastex7}
\usepackage{newtxtext,newtxmath}

\usepackage{graphicx}
\usepackage{lipsum}
\usepackage{caption}
\usepackage[T1]{fontenc}
\usepackage[utf8]{inputenc}
\usepackage{multirow}
\usepackage{natbib}
\usepackage{xcolor} 
\usepackage{booktabs}

\usepackage{amsmath} 

\graphicspath{ {./Figures/} }

\usepackage{float}
\usepackage{hyperref}

\usepackage{CJKutf8}

\begin{document}
\begin{CJK}{UTF8}{mj}
   
   \title{On the Mass Budget Problem of Protoplanetary Disks: Streaming Instability and Optically Thick Emission}

\correspondingauthor{Daniel Godines}

\author[0000-0001-8495-8205]{Daniel Godines}
\affiliation{Department of Astronomy, New Mexico State University, PO Box 30001, MSC 4500, Las Cruces, NM 88003-8001, USA}
\email[show]{godines@nmsu.edu}

\author[0000-0002-3768-7542]{Wladimir Lyra}
\affiliation{Department of Astronomy, New Mexico State University, PO Box 30001, MSC 4500, Las Cruces, NM 88003-8001, USA}
\email{wlyra@nmsu.edu}

\author[0000-0001-8123-2943]{Luca Ricci}
\affiliation{California State University, Northridge. Department of Physics and Astronomy, 18111 Nordhoff St, Northridge, CA 91330, USA}
\email{luca.ricci@csun.edu}

\author[orcid=0000-0003-2589-5034,gname=Chao-Chin,sname=Yang]{Chao-Chin Yang (楊朝欽)}
\affiliation{Department of Physics and Astronomy, The University of Alabama, Box 870324, Tuscaloosa, AL 35487-0324, USA}
\email{ccyang@ua.edu}

\author[0000-0002-3771-8054]{Jacob B. Simon}
\affiliation{Department of Physics and Astronomy, Iowa State University, Ames, IA 50010, USA}
\email{jbsimon@iastate.edu}

\author[orcid=0000-0003-2719-6640,gname=Jeonghoon,sname=Lim]{Jeonghoon Lim (임정훈)}
\affiliation{Department of Physics and Astronomy, Iowa State University, Ames, IA 50010, USA}
\email{jhlim@iastate.edu}

\author[0000-0001-6259-3575]{Daniel Carrera}
\affiliation{Department of Astronomy, New Mexico State University, PO Box 30001, MSC 4500, Las Cruces, NM 88003-8001, USA}
\email{carrera4@nmsu.edu}

%
%


\begin{abstract}
   {Statistical studies of protoplanetary disks and exoplanet populations often exhibit a ``missing mass'' problem, where observed dust masses in (sub-)millimeter surveys are significantly lower than expected when compared to the mass of evolved exoplanetary systems.}
   {We investigate how the streaming instability and subsequent planetesimal formation in protoplanetary disks might solve this missing mass problem when (sub-)millimeter observations are interpreted under the assumption of optically thin emission.}
   {We conduct hydrodynamical simulations of the streaming instability with self-gravity after which radiative transfer calculations with dust scattering are performed to measure the (sub-)millimeter intensity. The measured intensity is then used to estimate the disk mass under the assumption of optically thin emission and compared to the true mass in the simulation to calculate the observational bias via the mass excess.}
   {We find that the emission from overdense filaments that emerge due to the streaming instability are optically thick at (sub-)millimeter wavelengths, leading to mass excess factors of $\sim 2-7$, even when the optically thick fraction is low.}
   {
   }

\end{abstract}
   



\section{Introduction}
\label{intro}

Quantifying the amount of material in protoplanetary disks is essential for understanding the timescales and processes of planet formation. Multi-wavelength studies of disks around Class I/II Young Stellar Objects (YSOs) have constrained the size and radial distribution of dust grains \citep{Ricci_2010a,Ricci+10,Ricci_2010,Tazzari+15}, highlighting the utility of (sub-)millimeter (mm) observations for tracing dust in young disks. Modern facilities such as the Atacama Large Millimeter/submillimeter Array (ALMA) and the Very Large Array (VLA) have enabled dust mass estimates for hundreds of nearby disks \citep[e.g.,][]{Ansdell+16,Barenfeld_2016,Pascucci+16,Tychoniec+18,Andrews+18,Cieza+18}. However, these estimates often rely on the assumption of optically thin (sub-)mm emission, which may not hold throughout the disk. Indeed, comparisons with exoplanet demographics have revealed that many disks appear to lack sufficient dust mass to form the observed planetary systems. 

Several studies have quantified this mass budget discrepancy. \citet{Najita_and_Kenyon_2014} found that Class II disks typically lack the mass required to form massive exoplanets ($\sim5 - 10~M_\oplus$). Similarly, \citet{Ansdell+17} reported that only $\sim$10\% of disks in the $\sigma$ Orionis cluster have enough dust to reproduce the planets in the solar system. \citet{Manara_2018} found that the median dust mass in Lupus and Chamaeleon I disks was $\sim$25\% lower than the heavy-element mass in observed exoplanet systems, across stellar masses $\lesssim$2 $M_\odot$. More recently, \citet{Mulders+21} showed that when observational biases are accounted for, the mass in disks and exoplanets around solar-mass stars are of similar magnitude. These findings pose a challenge for planet formation theories, especially if pebble accretion is as inefficient as models predict, which typically capture only $\sim$1-10\% of the (sub-)mm-sized grains from the disk \citep{Guillot_2014}. If Class II dust reservoirs truly represent the progenitor material for giant planet cores, then accretion efficiencies closer to unity would be required. 

One proposed resolution is that planet formation occurs rapidly, with much of the dust already converted into planetesimals ($\gtrsim$~km-sized) by the Class II phase, rendering them invisible to (sub-)mm observations. Supporting this view, observations indicate a rapid decline in dust mass between the Class 0 and Class I/II phases \citep{Tychoniec+18,Tychoniec+20,Tobin+20}, with Class 0 disks appearing massive enough to reproduce known planetary systems \citep{Greaves+11,Najita_and_Kenyon_2014}. Nevertheless, planet formation during these early, potentially gravitationally unstable stages remains poorly understood.

Another plausible solution to the mass budget deficit problem is that protoplanetary disks may not be optically thin at (sub-)mm wavelengths. As \citet{Ricci_2012} showed, localized dust overdensities can create regions of high optical depth that can effectively ``hide'' mass from (sub-)mm observations. Moderate optical depths can arise from large dust column densities due to particle clumping driven by the streaming instability \citep{Scardoni+21,Rucska_2023,Scardoni+24}. This instability is energized by a velocity differential between the gas and dust, which resonates with inertial waves in the disk \citep{Squire+18}. The resulting streaming turbulence driven by the instability may ultimately lead to localized filaments of high dust concentration \citep{Yang_2014,Li+18, Schafer+24}. Even a few such regions can produce optically thick (sub-)mm emission, causing significant underestimation of disk dust masses when the optically thin assumption is applied.

The total solid mass in a disk may also be underestimated because a substantial fraction of dust mass could already be locked into planetesimals ($\lesssim10^2$~km-sized) or larger pebbles ($\gtrsim$10~cm), which are not effectively traced by radio observations. The streaming instability, long considered a leading mechanism for planetesimal formation \citep{Johansen+14}, plays a critical role in converting small dust grains into these larger bodies. Therefore, both optically thick emission and early planetesimal formation may contribute to the observed mass budget discrepancy, with masses derived from (sub-)mm fluxes serving as lower limits on the total solid content.

To address this discrepancy, it is essential to quantify how much mass may be concealed by optically thick (sub-)mm emission. \citet{Liu_2022} demonstrated that assuming optically thin emission at the 1.3 mm wavelength can lead to substantial underestimation of disk mass. While in extreme cases the mass can be underestimated by up to two orders of magnitude -- depending on disk properties such as dust mass, inclination, and internal substructure -- their fiducial model yielded a more modest mass underestimation of $\sim$1.4. \citet{Rucska_2023} conducted a similar analysis using streaming instability simulations with self-gravity and realistic grain size distributions, finding mass correction factors of order unity. However, their study did not account for scattering effects, which can further reduce the emergent flux and lead to additional bias in mass estimates \citep{Zhu_2019}. A more accurate analysis must therefore consider both opacity effects and mass hidden in planetesimals.

In this paper, we present numerical simulations of the streaming instability that include multiple grain sizes to investigate the observational consequences of optically thick dust emission and early planetesimal formation. We perform radiative transfer calculations to model the emergent (sub-)mm flux densities, incorporating both absorption and scattering effects. We then estimate the disk mass using the commonly adopted optically thin approximation, which relates the observed (sub-)mm flux to the dust column density. By comparing these inferred masses to the true dust mass in the simulation, we quantify the degree of underestimation and assess the bias introduced when the optically thin assumption does not hold.

This paper is organized as follows. In Section 2, we describe the simulation setup, the adopted disk model, and the radiative transfer calculations used to model the observable emission. In Section 3, we present the results of our comparison between the true and inferred disk masses. Section 4 discusses the implications of our findings, followed by concluding remarks in Section 5 on the limitations of the optically thin assumption and its impact on dust mass estimates in protoplanetary disks.

\section{Methods}
\label{sec:methods}

The simulations for this study were performed using the {\sc Pencil Code}, a high-order finite-difference code for astrophysical fluid dynamics \citep{Brandenburg_2002, Pencil_2021}. We utilized a shearing box approximation to model a local, co-rotating patch of the disk. In this framework, the equations of motion are solved in Cartesian coordinates using a local expansion of the Keplerian potential, which captures tidal forces, Coriolis effects, and the background shear \citep{Hawley_1995, Brandenburg_1995}. The curvature of the disk is neglected, and the dust is modeled as Lagrangian superparticles. 

We investigate the streaming instability with dust self-gravity enabled in order to quantify the impact of gravitationally collapsed planetesimals on the observed (sub-)mm emission. Including self-gravity is essential in this context, as dust concentrations driven by the streaming instability can exceed the critical threshold for gravitational collapse, leading to the formation of planetesimals. Once formed, these bodies no longer contribute significantly to (sub-)mm emission but can continue to grow via pebble accretion, thereby increasing the mass excess -- which we define as the difference between the true solid mass and that inferred from continuum observations. Unlike non-self-gravitating shearing box simulations, which are scale-free and can represent arbitrary disk locations, the inclusion of self-gravity breaks this scale invariance and anchors the simulation to a specific disk radius. Consequently, proper unit conversions are required to both initialize the simulation and to compute the radiative transfer accurately. To this end, we construct a disk model that provides the necessary physical parameters for a self-consistent treatment, as described in the following subsection.

\subsection{Disk Model}
\label{disk_model}

\begin{figure}[ht!]
  \includegraphics[width=\columnwidth]{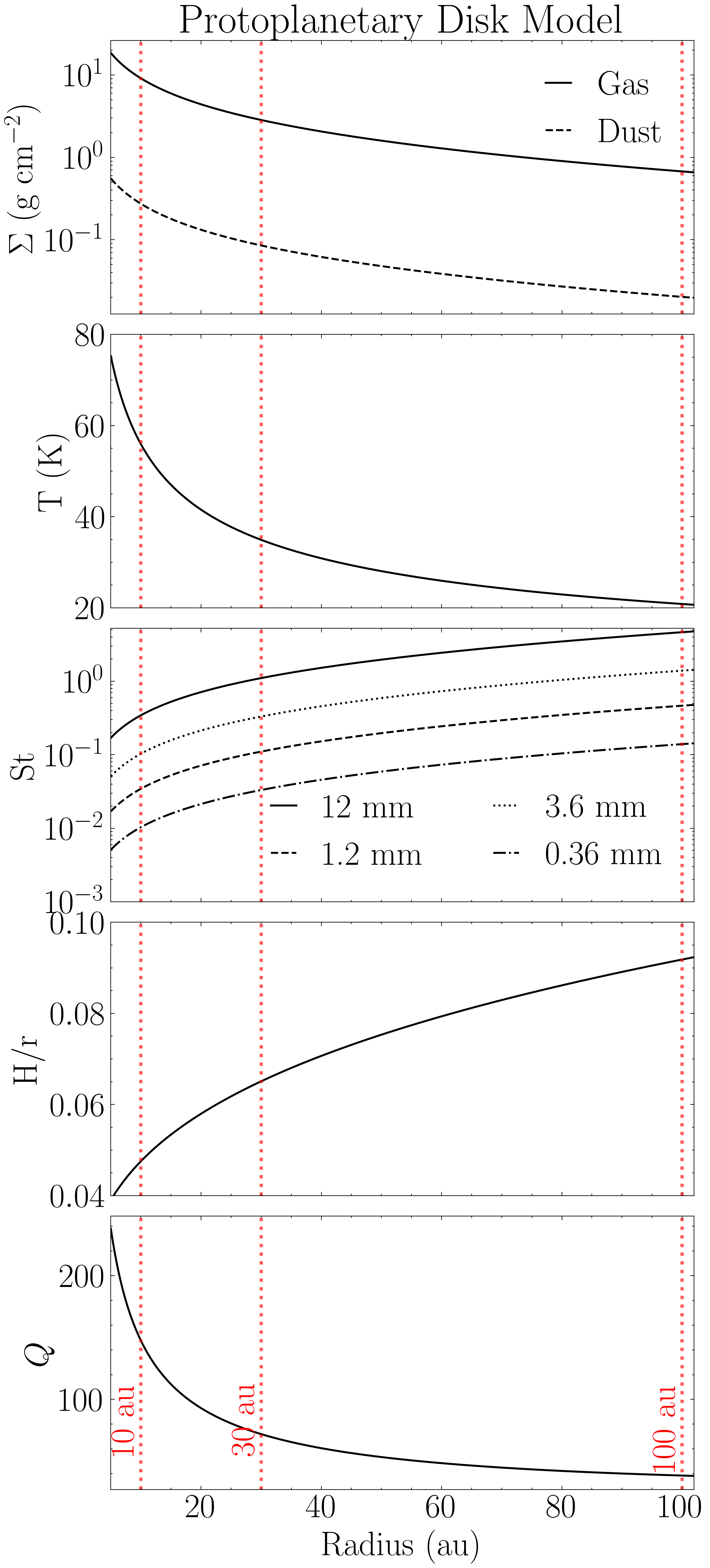}
  \caption{Radial profiles of key physical properties in our protoplanetary disk model. We model a disk orbiting a solar-mass star with a total mass of $M_{\rm disk} = 0.02~M_\odot$, composed of dust grains with an internal density of $\rho_\bullet = 1.675~\rm (g~cm^{-3})$. The vertical red dashed lines indicate the radial locations of our polydisperse, self-gravitating simulations. The simulations were set up with Stokes numbers corresponding to the location of the four particular grain sizes listed in the legend of the third panel.}
  \label{fig:disk_model_figure}
\end{figure}

Although we conduct three-dimensional, vertically stratified simulations, the setup is based on a vertically thin, axisymmetric disk with a total mass of $M_{\rm disk}=0.02~M_\odot$. This 2D model provides a mapping between radial location and gas surface density. The radial gas column density profile, $\Sigma_g (r)$, follows the self-similar solution of \citet{Lynden_1974},

\begin{equation}
\label{eq:sigma_g}
\Sigma_g(r) = \frac{M_{\rm disk}}{2\pi r_c^2} \left(\frac{r}{r_c}\right)^{-1} \exp\left(-\frac{r}{r_c}\right),
\end{equation}
where $r_c$ is the characteristic radius of the disk, which we set to 300~au. Assuming vertical isothermality, the midplane temperature is prescribed by a radial power-law profile,

\begin{equation}
\label{eq:temperature_profile}
T(r) = T_0 \left(\frac{r}{\rm au}\right)^{-q},
\end{equation}
where $T_0$ represents the temperature at $r=1$~au. We adopt $T_0=150$~K and $q=3/7$~K \citep{Chiang_1997}.

The pressure gradient is parameterized using the dimensionless $\Pi$ parameter \citep{Bai_2010}, 

\begin{equation}
\label{eq:big_pi}
\Pi \equiv \frac{\Delta v}{c_s},
\end{equation}
where $\Delta v$ is the azimuthal velocity difference between the gas and dust, and $c_s$ is the isothermal sound speed. The sound speed is related with the temperature by

\begin{equation}
\label{eq:sound_speed}
c_s = \sqrt{\frac{\gamma T k_B}{\mu m_H}},
\end{equation}
where $\gamma$ is the adiabatic index and $\mu$ is the mean molecular weight of the gas, $k_B$ is the Boltzmann constant, and $m_H$ the mass of a hydrogen atom. We adopt $\gamma = 1$ to represent efficient radiative cooling and set $\mu = 2.3$ for a 5 parts molecular hydrogen and 2 parts helium mixture. The gas scale height of the disk is then given by,

\begin{equation}
\label{eq:scale_height}
H \equiv \frac{c_s}{\Omega_K},
\end{equation}
where $\Omega_K$ is the Keplerian angular frequency. With the assumed vertical isothermality, the midplane gas volume density is given by $\rho_g=\Sigma_g(r)\left(\sqrt{2 \pi}H\right)^{-1}$.

Dust grain sizes are characterized by their Stokes number, $\mathrm{St}$, which quantifies the timescale over which solid particles couple to the surrounding gas,

\begin{equation}
\label{eq:stokes_number}
    \mathrm{St} \equiv \Omega_K t_s,
\end{equation}
where $t_s$ is the stopping time of the solid particles. 

In the Epstein regime, valid for mm-sized dust grains in typical disk conditions, the Stokes number at the disk midplane is given by 

\begin{equation}
\label{eq:stoke}
    \text{St} = \frac{\pi}{2} \frac{\rho_\bullet a}{\Sigma_g (r) },
\end{equation}
where $\rho_\bullet$ and $a$ are the internal density and radius of the dust grains, respectively \citep{epstein1924resistance, Weidenschilling_1977}. We adopt $\rho_\bullet=1.675~(\rm g \rm~cm^{-3})$, based on the dust composition model presented in the Disk Substructures at High Angular Resolution Project \citep[DSHARP;][]{Birnstiel_2018}, which includes refractory organic materials in addition to astronomical silicates, water ice, and troilite. We adopt the DSHARP dust model to extract representative absorption and scattering opacity coefficients for the grain sizes in our simulations (see Section~\ref{opacities}).
  
To model planetesimal formation via gravitational collapse of dust overdensities, we solve Poisson’s equation for the dust potential, 

\begin{equation}
    \nabla^2 \Phi = 4\pi G \rho_d,
    \label{eq:Poisson}
\end{equation}
where $G$ is the gravitational constant, and $\Phi$ and $\rho_d$ are the gravitational potential and dust mass volume density, respectively. This equation is solved using Fast Fourier Transforms following the approach of \citet{Johansen+2007}. The gravitational stability of the disk is described by the Toomre $Q$ parameter \citep{Safronov1960, Toomre_1964},

\begin{equation}
\label{eq:toomre_q}
    Q = \frac{c_s \Omega_K}{\pi G \Sigma_g},
\end{equation}
and the strength of local self-gravity is expressed using the dimensionless parameter $\tilde{G}$, 

\begin{equation}
\label{eq:tilde_g}
    \tilde{G} = \frac{4 \pi G \rho_d}{\Omega_K^2} \equiv \sqrt{\frac{8}{\pi}} Q^{-1}.
\end{equation}

Fig.~\ref{fig:disk_model_figure} presents the radial profiles of the key disk parameters, including gas \& dust surface density, temperature, Stokes numbers, disk aspect ratio, and Toomre Q. We evaluate self-gravity at three representative disk radii: 10, 30, and 100 au. In our simulations, we include four dust species with radii of 12, 3.6, 1.2, and 0.36 mm. As shown by \citet{Krapp_2019}, collective multi-species effects in unstratified simulations only arise for dust-to-gas ratios $\gtrsim 10^{-2}$ when the size distribution is resolved with at least 16 – 32 discrete species. Our four-species simulations fall below this threshold, therefore the species exhibit no collective behavior and evolve largely independently. The third panel of Fig.~\ref{fig:disk_model_figure} shows the variation in Stokes number across the disk for each species, although each simulation is conducted at a fixed radial location as per the density constraints set by the Stoke numbers \& $Q$ value. The central positions of the shearing boxes are marked by red dashed lines in all panels. The following subsection outlines the setup and gravitational constraints of the three simulations conducted in this study.

\subsubsection{Simulations}
\label{simulations}

\begin{figure}
\centering
\includegraphics[width=\columnwidth]{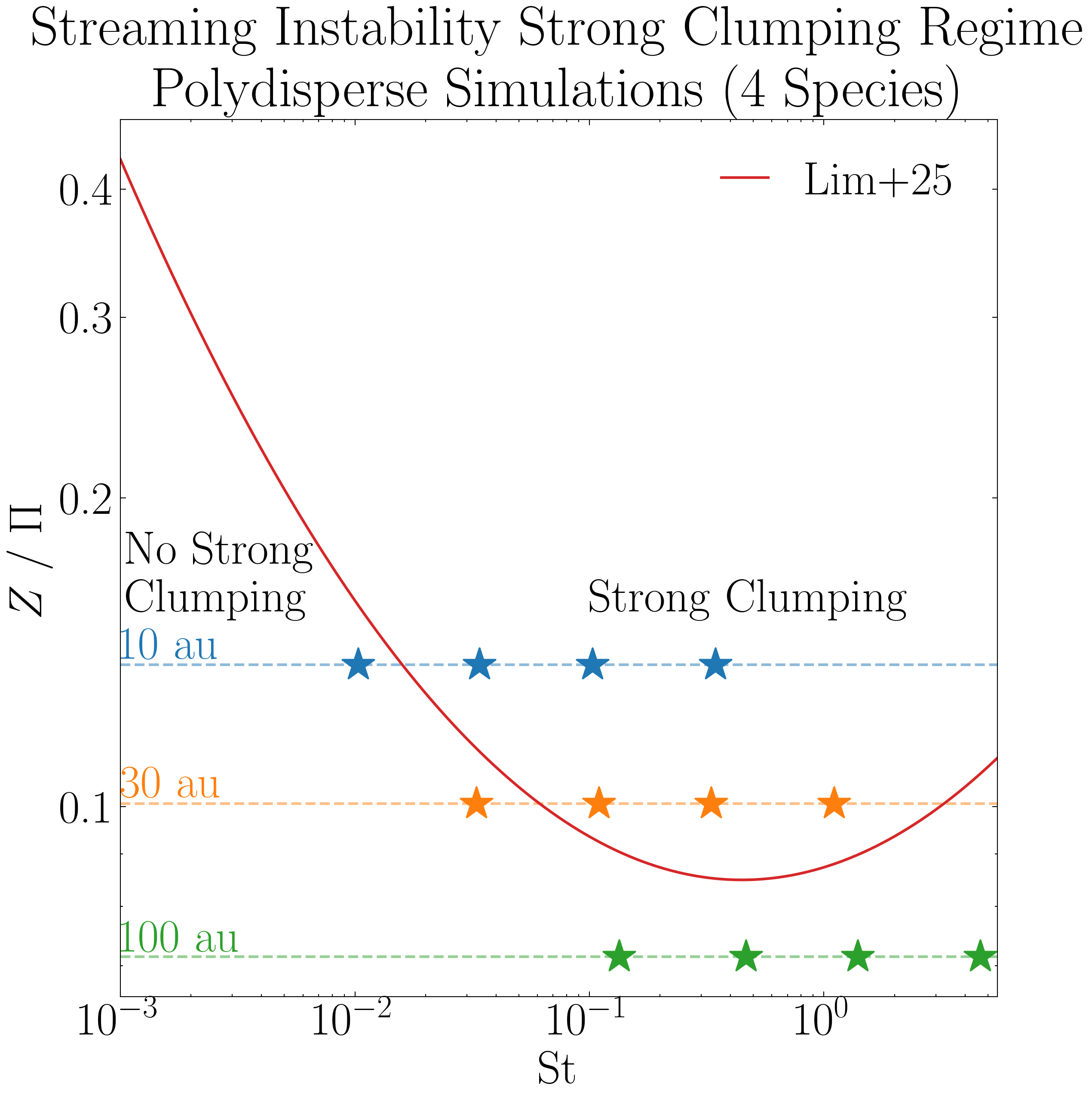}
\caption{Regime in which the streaming instability induces strong particle clumping. The normalized solid-to-gas ratio of each simulation is shown on the y-axis and is demarcated by distinct dashed lines. The corresponding stars represent the four species in each simulation, which are placed within the clumping boundary according to the respective Stokes number.}
\label{fig:si_criteria}
\end{figure}

We conducted three high-resolution ($256^3$) streaming instability simulations using the {\sc Pencil Code} to evaluate the mass excess at three different radial locations in the disk. The simulations were performed at 10, 30, and 100~au, each within a simulation domain of $L_x=L_y=L_z=0.2H$, populated with $10^6$ superparticles and evolved for 100 orbital periods ($P$). Each simulation was set up with four bins of uniform dust distribution consisting of four species with grain sizes of 12, 3.6, 1.2, and 0.36 mm (see third panel of Fig.~\ref{fig:disk_model_figure}). Each simulation adopts a different pressure gradient parameter $\Pi$, computed from the local disk conditions at its respective radial location. Table~\ref{tab:simulations} summarizes the key simulation parameters, including the strength of self-gravity and the $\Pi$ values corresponding to the three disk locations.

All three simulations are initialized with a solid-to-gas ratio of $Z=0.03$, defined as 

\begin{equation}
\label{eq:solid_to_gas}
    Z \equiv \frac{\Sigma_{d,0}}{\Sigma_{g,0}},
\end{equation}
where $\Sigma_{d,0}$ and $\Sigma_{g,0}$ are the vertically integrated initial column densities of the dust and gas, respectively. In this work, we treat the solid-abundance-to-headwind criterion, $Z/\Pi$, as a critical parameter for particle clumping, although we note that the precise threshold for strong clumping remains an active area of research \citep{Johansen+2009, Carrera+15, Yang_2017, Li_2021, Lim+25}. Since our simulations include four dust species that evolve relatively independently with other species, we normalize this ratio by the number of species, effectively using $Z/\Pi/4$ as the relevant quantity. Fig.~\ref{fig:si_criteria} shows the positions of our three simulations in this normalized clumping regime, as identified by \citet{Lim+25}, who studied 2D axisymmetric streaming instability in the absence of external turbulence.

Fig.~\ref{fig:si_criteria} shows that, after normalization by the number of species, the 10 and 30~au simulations fall within the strong clumping regime for the three largest grain sizes. In contrast, the simulation at 100~au lies below the clumping threshold and serves effectively as a control case, where the absence of strong clumping and the low dust surface density ($\Sigma_d\lesssim1~\rm~g\rm~cm^2$; top panel of Fig.~\ref{fig:disk_model_figure}) results in optically thin conditions throughout the  domain. A commonly used criterion to determine when dust clumps become gravitationally bound is the Hill density, $\rho_H$ (also known as the Roche density),

\begin{equation}
\label{eq:hill_density}
    \rho_H \equiv \frac{9 \Omega_K^2}{4\pi G}.
\end{equation}
This is the critical density above which self-gravity overcomes tidal shear. While $\rho_H$ does not account for turbulent diffusion, we adopt it as a practical collapse threshold, and it thus serves as an effective criterion that also accounts for the local turbulence in our simulations. As shown by (e.g., \citealt{Johansen+12,Johansen+15_SciA}), overdensities generated by the streaming instability can collapse under their own self-gravity if the local dust density exceeds $\rho_H$. Furthermore, \citet{Klahr+20} demonstrated that turbulence driven by the streaming instability imposes a characteristic minimum mass for gravitational collapse, effectively setting a lower bound on planetesimal formation. Small-scale pebble clouds are rapidly diffused before they can collapse, whereas sufficiently large overdensities remain gravitationally bound and form planetesimals. In our simulations we account for this effect by enabling self-gravity after two orbits and allowing dust clumps to overcome turbulent stirring and collapse into ``sink particles'' only if their density reaches $\geq2\rho_H$. This more stringent criterion ensures that only the most overdense regions undergo gravitational collapse, filtering out small-scale overdensties that would otherwise be disrupted by turbulent diffusion. 

\begin{table}
  \centering
  \caption{Polydisperse simulations with self-gravity. The Stokes numbers at each location correspond to grain sizes of radii 12, 3.6, 1.2, and 0.36 mm, respectively.}
  \label{tab:simulations}
  \begin{tabular}{cccc}
    \toprule
    Radius (au) & Stokes Numbers & $\Pi$ & $\tilde{G}$ \\
    \midrule
    10 & 0.345, 0.103, 0.034, 0.0103 & 0.0545 & 0.010 \\
    30 & 1.105, 0.331, 0.110, 0.033 & 0.0745 & 0.022 \\
    100 & 4.651, 1.395, 0.465, 0.134 & 0.1050 & 0.042 \\
    \bottomrule
  \end{tabular}
\end{table}

\begin{figure*}[ht!]
\includegraphics[width=\textwidth]{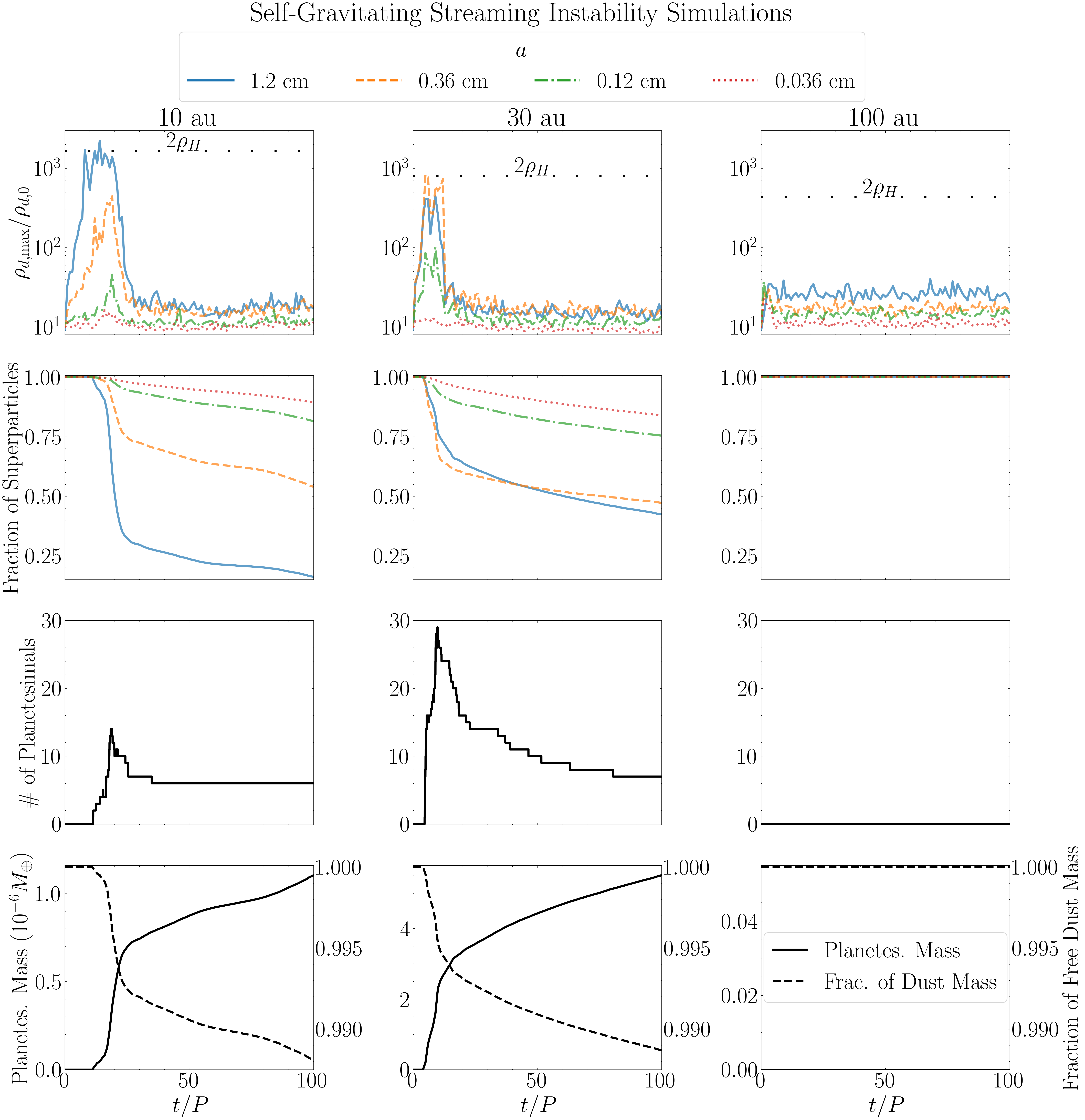}
    \caption{The temporal evolution of three polydisperse streaming instability simulations with self-gravity, shown column-wise. The top row presents the maximum dust density over time, alongside the modified Hill density criterion for planetesimal formation. Four lines represent the evolution of the different species, with the second row showing the quantity of each species that is present over time (i.e., not accreted by planetesimals). The third row displays the number of planetesimals present in the simulation, and which decrease over time due to mergers. The bottom row shows the total dust mass that these planetesimals lock up, while the twin y-axis shows the corresponding mass fraction that is available.}
    \label{fig:simulations}
\end{figure*}

The dust evolution over time is portrayed in Fig.~\ref{fig:simulations}, with the columns separating the three different locations of our simulations. The top row portrays the maximum dust density, broken down by dust species, which are represented by the four unique line styles. In this plot we also denote the Hill density threshold we adopt in each case, which shows how at 10~au it is the $\sim$cm-sized grains (St~$\sim$~0.3) that reach the critical densities and become planetesimals, while in the simulation at 30~au the most active dust grains are those of 0.36~cm size but with a similar Stokes number of St$\sim$0.3. On the contrary, the dust densities in the 100~au case are relatively constant with $\rho_{d, \rm max}$ reaching $\sim$10\% of the required critical density. The second row displays the fraction of superparticles, per species, that is present in the simulation over time (i.e., not yet locked up in the growing planetesimals). 

Dust grains in the simulations are merged into sink particles when they reside within a single grid cell. As planetesimals grow and accrete mass, they continue to merge, resulting in a declining number of bodies during and after the initial phase of rapid planetesimal formation. This evolution reflects the effective accretion rate as a function of dust species, with larger, more actively accreting grains decreasing by roughly an order of magnitude, while smaller grains remain tightly coupled to the gas \citep{Yang_Zhu_21,Canas+24}. Although this merging is a numerical treatment that conserves mass and energy, it does not resolve detailed collision physics. Nevertheless, this behavior illustrates that in real disks, planetesimal and pebble accretion can hide substantial mass within compact, optically thick regions that are effectively invisible in (sub-)mm observations. As no planetesimal formation occurs at 100~au, the dust grain population remains unchanged.

The third row of Fig.~\ref{fig:simulations} illustrates the number of planetesimals that develop over time. Soon after midplane settling, the formation rate is high and the planetesimals acquire the majority of their mass. This initial burst of planetesimal formation is observed at $t/P \sim 20-30$, during which $\sim 1\%$ of the available dust mass is converted into planetesimals. By the end of the run, the number of planetesimals stabilizes at six for the simulation at 10 au and seven for the 30 au case. The fourth row illustrates the mass of the growing planetesimals, while the twin y-axis presents the corresponding decrease in the dust mass present in the simulation. The planetesimals that form at 10 au lock up $\sim 10^{-6}\,M_\oplus$ in dust mass, while the simulation at 30 au accretes $\sim 5.5 \times 10^{-6}\,M_\oplus$, respectively. The available dust mass remains constant at 100 au, as this simulation produces no planetesimals. In the following subsection, we review how the opacities are calculated for these polydisperse simulations and how the subsequent optical depth maps are generated.

\subsection{Optical Depth \& Opacities}
\label{opacities}

To calculate the optical depth we assume that the disk, as observed from Earth, is oriented face-on. This restriction allows us to perform one-dimensional radiative transfer calculations at the disadvantage of not being able to quantify the mass excess as a function of disk inclination. We note however, that higher disk inclination yields larger optical depths (see e.g., \citealt{Liu_2022}), therefore this face-on orientation gives a lower limit. The effective optical depth along the $z$-direction can be computed with
\begin{equation}
\label{eq:optical_depth}
\tau_\nu^{\rm eff} = \kappa_\nu^{\rm eff} \Sigma_d,
\end{equation}
where $\Sigma_d$ is the dust column density and $\kappa_\nu^{\rm eff}$ is the effective monochromatic dust opacity which accounts for both absorption and scattering effects. This effective opacity is defined as the sum of the absorption opacity coefficient, $\kappa_\nu$, and the corresponding scattering opacity, $\sigma_\nu$, which are both a function of the grain size, $a$,

\begin{equation}
\label{eq:kappa_eff}
    \kappa_\nu^{\rm eff} \equiv \kappa_\nu (a) + \sigma_\nu (a).
\end{equation}

\begin{figure}[ht!]
  \centering
    \centering
\includegraphics[width=\columnwidth]{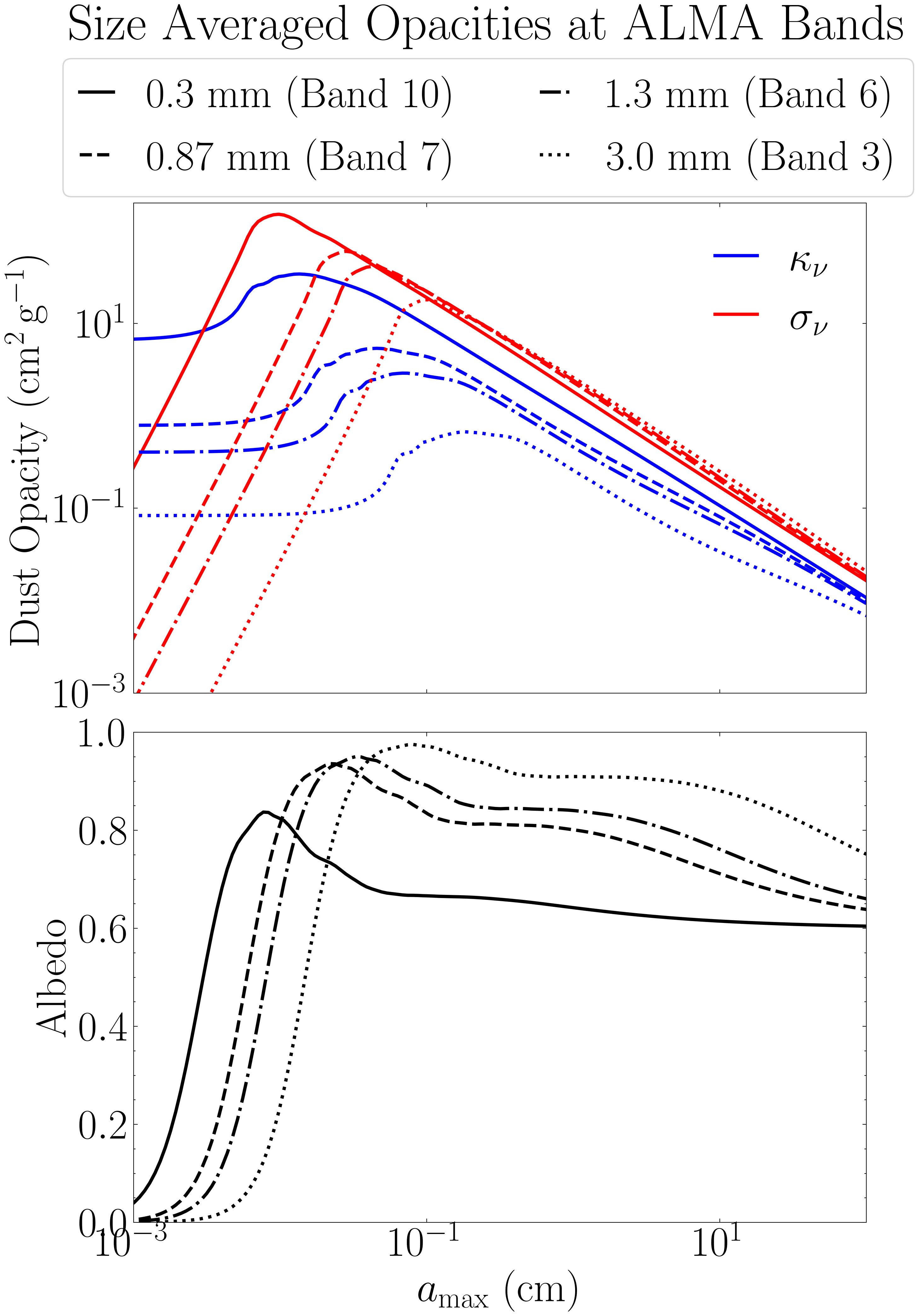}
    \caption{Dust opacity and albedo across ALMA bands. The top panel shows the absorption (blue) and scattering (red) opacities from the DSHARP dust model for four different ALMA bands. The second panel presents the corresponding albedos.}
\label{fig:opacities_and_albedo_1}
\end{figure}

\begin{figure}[ht!]
  \centering
    \centering
\includegraphics[width=\columnwidth]{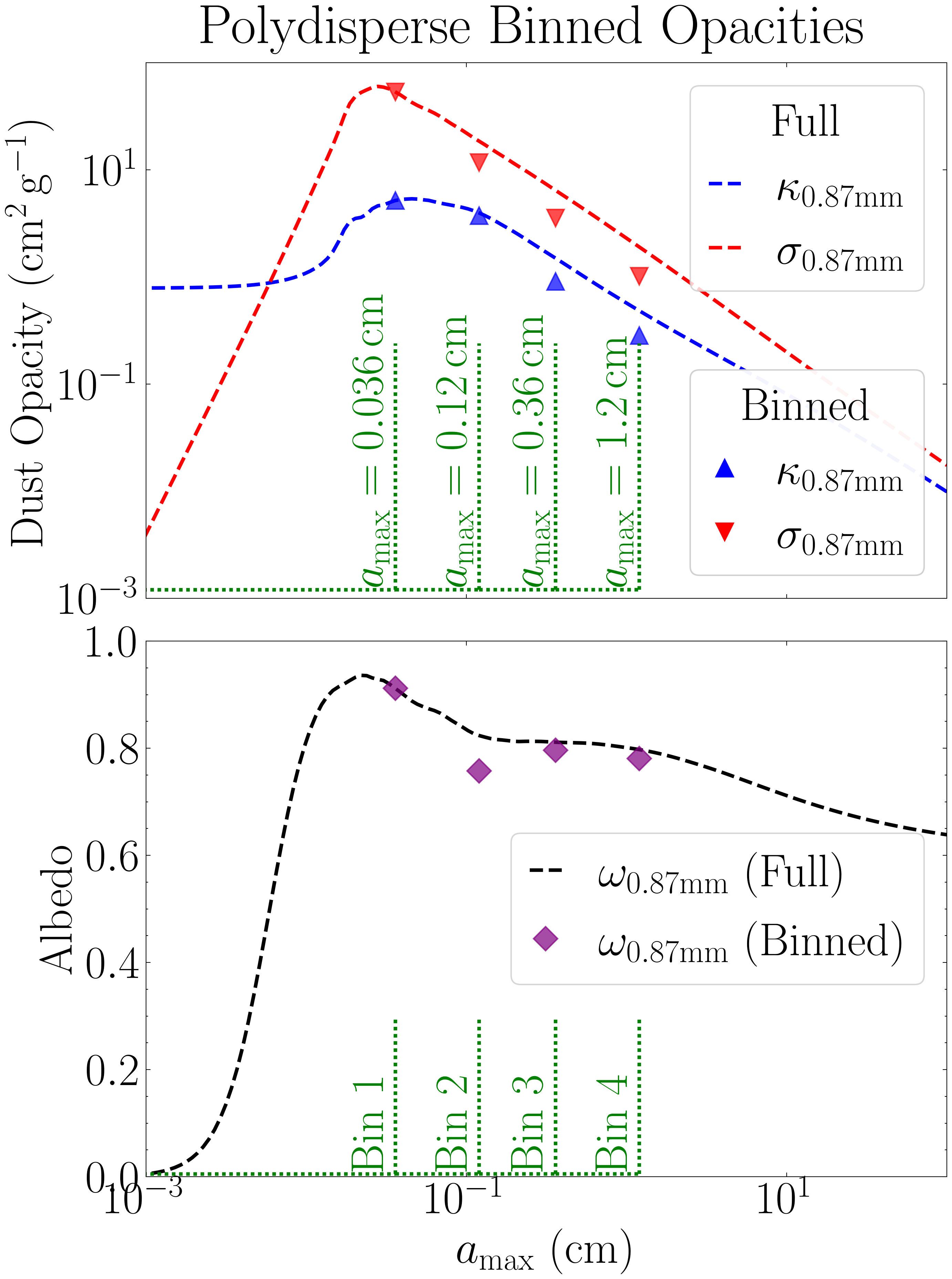}
    \caption{Grain size binning for multi-species analysis. The top panel shows the absorption (blue) and scattering (red) opacities for ALMA Band 7. Dashed lines represent the full grain size distribution, while colored markers indicate the binned values used in our radiative transfer calculations. The bottom panel presents the corresponding albedos.}
\label{fig:opacities_and_albedo_2}
\end{figure}

Calculating representative dust opacity coefficients for varying grain sizes in protoplanetary disks is complicated by several factors, including the assumed material composition, grain morphology, and disk temperature. In this work, we utilize results from the DSHARP project \citep{Birnstiel_2018} to determine the dust opacity coefficients. Using the DSHARP dust model, we estimate the particle size-averaged absorption and scattering opacities at the (sub-)mm wavelengths observed by ALMA. The dust opacities depend on the grain size distribution, which for truncated power-laws is parameterized by a power-law index, $p$, as well as the minimum and maximum grain sizes, $a_{\rm min}$ and $a_{\rm max}$, respectively. Observations suggest that the grain size distribution in protoplanetary disks follows an inverse power-law of the form $n(a)\,da \propto a^{-p}\,da$ \citep{Mathis+1977}, spanning sizes from (sub-)mm to cm scales (see \citealt{Birnstiel+11, Testi_2014}, and references therein). This treatment has been widely applied to interpret (sub-)mm observations of young disks in nearby star-forming regions \citep{Ricci_2010, Ricci_2010a, Testi_2014} and is thus adopted in this work.

Before we consider our models of four distinct dust species instead of a continuous size distribution, we first reproduce the DSHARP opacities as follows. We assume a power-law index of $p=2.5$ \citep{DAlessio+01}, with $a_{\rm min}=10^{-5}$~cm fixed as per the DSHARP dust model. Given a dust grain mass density distribution, $n(a) m(a)$, the effective opacity can be calculated as \citep{Miyake_1993},

\begin{equation}
\label{opacity_eqn}
    \kappa_\nu^{\text{eff}} = \frac{\int_{a_{\text{min}}}^{a_{\text{max}}} n(a)m(a) \ [ \kappa_\nu(a) + \sigma_\nu(a) ] \, da}{\int_{a_{\text{min}}}^{a_{\text{max}}} n(a)m(a) \, da}.
\end{equation}
Using Equation~\ref{opacity_eqn}, we calculate the absorption and scattering opacities for a range of grain sizes, as well as the corresponding single scattering albedos,

\begin{equation}
\label{albedo_eqn}
\omega_\nu \equiv \frac{\sigma_\nu}{\kappa_\nu + \sigma_\nu}.
\end{equation}
These opacities are computed at four of the (sub-)mm bands ALMA observes: 3 mm (Band 3), 1.3 mm (Band 6), 0.87 mm (Band 7), 0.3 mm (Band 10). The dust opacities at these frequencies are displayed in the top panel of Fig.~\ref{fig:opacities_and_albedo_1} for a range of maximum grain sizes ($10^{-3}-10^2$~cm). The four line styles differentiate between the different ALMA bands, with the absorption opacities shown in blue and the scattering opacities in red. The (sub-)mm opacities decrease with wavelength, with the scattering coefficients being approximately an order of magnitude larger than the absorption opacities. The lower panel shows the corresponding albedos which increase with wavelength. The high (sub-)mm scattering opacities at these grain sizes yields high albedo values of $\omega_\nu\sim0.6-0.95$.

We now consider the opacities for our models of four distinct dust species. In the polydisperse case, the effective opacity is computed at each grid cell by averaging over multiple species, each with distinct sizes and corresponding opacities. This is done using a density-weighted average,

\begin{equation}
\label{opacity_eqn_discrete}
    \kappa_\nu^{\text{eff}} = \frac{\sum\limits_{i=1}^{n} \rho_d(a)_i \left[ \kappa_\nu(a)_i + \sigma_\nu(a)_i \right]}{\sum\limits_{i=1}^{n} \rho_d(a)_i},
\end{equation}
where $\kappa_\nu(a)_i$ and $\sigma_\nu(a)_i$ are the absorption and scattering opacities for dust species, $i$, $\rho_d(a)_i$ is the corresponding dust density, and $n=4$ is the number species in our simulation.

When computing the effective opacities, it is important to avoid double-counting the smaller grains, which would happen if the lower bound of every size bin were fixed at $a_{\rm min}=10^{-5}$~cm. If the minimum grain size in the distribution is fixed, the small end of the distribution would be overrepresented in the integrals for Equation~\ref{opacity_eqn}, artificially inflating the contribution from high-opacity, small grains. To avoid this, we construct non-overlapping grain size bins. Each dust species in the simulation represents the maximum dust size ($a_{\rm max}$) of one bin, while the minimum dust size ($a_{\rm min}$) of that bin is slightly larger than the next smaller dust species. For the smallest species, we assume $a_{\rm min}=10^{-5}$~cm. Equation~\ref{opacity_eqn_discrete} can then be used to average the opacities, and the binning ensures that each grain size contributes only once to the opacity calculation.

Fig.~\ref{fig:opacities_and_albedo_2} illustrates this binning approach for ALMA Band 7 (0.87~mm). The top panel shows the absorption (blue) and scattering (red) opacities, while the bottom panel presents the corresponding albedos. Dashed lines show the results from the full grain size distribution, identical to those in Fig.~\ref{fig:opacities_and_albedo_1}, while the scatter points represent the opacities computed from the binned distributions. The first bin spans grain sizes from $10^{-5}$~cm to 0.036~cm (`Bin 1’), corresponding to the smallest dust species in our simulations. Since this bin encompasses the full lower end of the DSHARP distribution, the binned opacity matches the value from the full distribution. For the remaining three bins, which correspond to the larger dust species, the binned opacities are systematically lower than those from the continuous distribution. This reflects the reduced contribution from the smaller grains in each bin, leading to more representative opacities when considering multi-species simulations.

\subsection{Radiative Transfer}
\label{radiative_transfer}

To determine the emergent intensity at the output plane where $z=L_z/2$, we solve the radiative transfer equation,
\begin{equation}
\frac{dI_\nu}{d\tau_\nu^{\rm eff}} = -I_\nu + S_\nu^{\rm eff},
\end{equation}
where $S_\nu^{\rm eff}$ is the effective source function that considers both absorption and scattering. The infinitesimal $d\tau_\nu^{\rm eff}$ is the change in effective optical depth, which represents how much the optical depth varies over an infinitesimal vertical distance, $dz$, along the vertical line of sight,

\begin{equation}
\label{dt_eff_eqn}
    d\tau_\nu^{\rm eff} = \kappa_\nu^{\rm eff} \rho_d dz.
\end{equation}
To calculate the outgoing intensity, we evaluate the Planck function for temperature-dependent blackbody radiation, $B_\nu\left(T\right)$, based on the temperature profile we adopt in this work (Equation~\ref{eq:temperature_profile}),

\begin{equation}
\label{eq:blackbody}
B_\nu\left(T\right) = \frac{2h\nu^3}{c^2} \frac{1}{\exp(h \nu / k_B T) - 1}.
\end{equation}
In conjunction with the effective emissivity, $j_\nu^{\rm eff}\equiv\kappa_\nu B_\nu+\sigma_\nu J_\nu$, the effective source function is defined as

\begin{equation}
    S_\nu^{\rm eff} = \frac{j_\nu^{\rm eff}}{\kappa_\nu^{\rm eff}} \equiv \frac{\kappa_\nu B_\nu + \sigma_\nu J_\nu}{\kappa_\nu + \sigma_\nu},
\end{equation}
where $J_\nu$ is the zeroth moment directional average of the intensity, $I_\nu$. 
This effective source function can be expressed in terms of the albedo as \citep{Mihalas_1978},

\begin{equation}
\label{eq:s_eff}
    S_\nu^{\rm eff} = \omega_\nu J_\nu + \left(1 - \omega_\nu\right) B_\nu.
\end{equation}
To calculate $J_\nu$ we used the analytical solution by \citet{Miyake_1993}, which assumes a constant disk temperature and neglects any incoming radiation at the disk surfaces,

\begin{equation}
\label{eq:j_nu_equation}
    \frac{J_\nu\left(\tau_d\right)}{B_\nu\left(T\right)} = 1 - \frac{\exp(-\sqrt{3\epsilon}~\tau) + \exp(\sqrt{3\epsilon}\left[\tau - \tau_d\right])}{\exp(-\sqrt{3\epsilon}~\tau_d)(1 - \sqrt{\epsilon}) + (\sqrt{\epsilon} + 1) },
\end{equation}
where we use the substitution, $\epsilon=1-\omega_\nu$. The optical depth at the disk surface is set to $\tau=0$, while the characteristic depth, given by $\tau_d=2\cos(\theta)/3$, follows the Eddington-Barbier relation, which states that the emergent intensity along a slanted ray is approximately equal to the source function at this depth. Here, $\theta$ represents the angle between the slanted rays and the $z$-direction. For the slab solution with scattering, the emergent rays originate from $\cos(\theta)=1/\sqrt{3}$ at all inclinations. Although this omits stellar irradiation, the approximation is consistent with our assumption of vertical radiative equilibrium in the absence of external sources. 

The general solution to the radiative transfer equation is given by

\begin{equation}
    \label{eq:rad_transfer}
    \begin{split}
    I_\nu (L_z/2) = I_\nu(-L_z/2)\exp(-\tau_\nu^{\rm eff}) \\
    & \hspace{-6em} + \int_{-L_z/2}^{L_z/2} j_\nu^{\rm eff}(z) \rho_d(z) \exp(-\left[ \tau_\nu^{\rm eff} - t_\nu (z) \right]) \ dz,
    \end{split}
\end{equation}
The first term represents the extinction of the original intensity, which is zero in the context of thermal dust emission in disks in which there is no back illumination. The second term in the solution accounts for the contribution from the dust emission along the entire column, incorporating two distinct optical depths. The depth-dependent optical depth, $t_\nu(z)$, represents the cumulative absorption and scattering by dust grains up to a point $z$ as the emission traverses the column,

\begin{equation}
     t_\nu(z) = \int_{-L_z/2}^{z} \kappa_\nu^{\rm eff}(z') \rho_d(z') dz'.
\end{equation} 
while the total optical depth, $\tau_\nu$, is $t_\nu(z)$ but evaluated only at the exit plane where $z=L_z/2$. 

Using Equation~\ref{eq:rad_transfer}, we construct a 2-dimensional intensity map (see Section~\ref{sec:results}), from which we quantify the filling factor,

\begin{equation}
\label{eq:ff_fraction}
f_{\rm fill, \nu} = \frac{N(\tau_\nu^{\rm eff} \geq 1)}{N_{\rm total}},
\end{equation}
where $N(\tau_\nu \geq 1)$ is the number of ($x$,~$y$) columns with optical depth above unity and $N_{\rm total}$ is the total number of columns in the map; and where the subscript ``$\nu$'' is used to indicate that this quantity is determined based solely on the flux densities at a particular band. We also define the optically thick fraction, $f_{\rm thick, \nu}$, which is the ratio of the output intensity to the intensity expected from an ideal blackbody at the same temperature (Equation~\ref{eq:blackbody}),

\begin{equation}
\label{eq:ff_thick}
    f_{\rm thick, \nu} = \frac{\langle I_\nu \rangle}{B_\nu}.
\end{equation}
Here, $\langle I_\nu \rangle$ is the mean of the intensity map so as to represent an unresolved pixel as would be observed by ALMA. By calculating the optically thick fraction under different contexts, we can quantify deviations from the optically thin assumption and in turn the observational bias that arises from localized high-density regions such as dust filaments and/or forming planetesimals.

\begin{table}
\centering
\begin{tabular}{c c c}
\hline
\textbf{Parameter} & \textbf{Equation} & \textbf{Note} \\
\hline
$\rho_\bullet$ &  &  Dust grain internal density \\
$\kappa_\nu$ &  & Absorption Opacity \\
$\sigma_\nu$ &  & Scattering Opacity \\
$\Sigma_g(r)$ & (\ref{eq:sigma_g}) & Gas surface density profile \\
$T(r)$ & (\ref{eq:temperature_profile}) & Temperature profile \\
$\Pi$ & (\ref{eq:big_pi}) & Pressure gradient \\
$c_s$ & (\ref{eq:sound_speed}) & Sound speed \\
$H$ & (\ref{eq:scale_height}) & Gas scale height \\
$\mathrm{St}$ & (\ref{eq:stokes_number}) & Stokes number \\
$Q$ & (\ref{eq:toomre_q}) & Toomre Q \\
$\tilde{G}$ & (\ref{eq:tilde_g}) & Self-gravity strength \\
$Z$ & (\ref{eq:solid_to_gas}) & Dust-to-gas ratio \\
$\rho_H$ & (\ref{eq:hill_density}) & Hill density \\
$\tau_\nu^{\rm eff}$ & (\ref{eq:optical_depth}) & Effective Optical Depth \\
$\kappa_\nu^{\rm eff}$ & (\ref{eq:kappa_eff}) & Effective Opacity \\
$\omega_\nu$ & (\ref{albedo_eqn}) & Single scattering albedo \\
$B_\nu(T)$ & (\ref{eq:blackbody}) & Planck function \\
$S_\nu^{\rm eff}$ & (\ref{eq:s_eff}) & Effective source function \\
$f_{\rm fill, \nu}$ & (\ref{eq:ff_fraction}) & Filling Factor \\
$f_{\rm thick, \nu}$ & (\ref{eq:ff_thick}) & Optically thick fraction \\
$\Lambda_{\nu}$ & (\ref{eq:me_equation}) & Mass excess \\
\hline
\end{tabular}
\caption{List of parameters and equations used in this study.}
\label{tab:parameter_table}
\end{table}

\subsection{Mass Excess}
\label{sec:mass_excess_calculation}

In protoplanetary disks, the emergent intensity from optically thin (sub-)mm emission ($\tau_\nu^{\rm eff}\ll1$) scales with the dust surface density, temperature, and the absorption opacity of the dust grains \citep{Hildebrand1983},

\begin{equation}
\label{eq:sigma_d_eqn}
\Sigma_{d, \rm obs} = \frac{\langle I_\nu\rangle}{B_\nu \kappa_\nu}.
\end{equation}
Here, $\Sigma_{d, \rm obs}$ defines the inferred surface density, estimated from the output intensity, $\langle I_\nu \rangle$.

This expression assumes an isothermal slab and neglects scattering, such that the source function is taken to be the Planck function (Equation~\ref{eq:blackbody}). This is valid in the optically thin limit, where most photons escape without interacting and in which scattering has minimal impact on the emergent intensity. However, this approximation breaks down in optically thick regions or when scattering opacities are comparable to or larger than absorption opacities, as is the case at (sub-)mm wavelengths for large grains (Fig.~\ref{fig:opacities_and_albedo_1}). 
In such cases, the effective source function (Equation~\ref{eq:j_nu_equation}) becomes vertically dependent, as scattering can redirect photons out of the line of sight. This introduces a potential bias in Equation~\ref{eq:sigma_d_eqn}, particularly near the optically thick-thin transition, where the approximation may underestimate the true dust surface density. While we treat scattering explicitly in the radiative transfer, surface density estimates are computed using this expression to quantify the associated observational bias.

Equation~\ref{eq:sigma_d_eqn} enables an estimate of the dust mass in a local region of the disk from the observed (sub-)mm flux densities, which we denote as $m_{\rm obs}$. In the context of our simulations, this corresponds to the product of $\Sigma_{d, \rm obs}$ and the simulation domain area. To quantify the bias introduced by the optically thin approximation, we define the mass excess as the ratio of the true dust mass in the simulation ($m_{\rm true}$) to the observed mass derived from Equation~\ref{eq:sigma_d_eqn}. Following the convention set forth by \citet{Liu_2022}, we denote the mass excess as $\Lambda_{\nu}$,

\begin{equation}
\label{eq:me_equation}
    \Lambda_{\nu} \equiv \frac{m_{\rm true}}{m_{\rm obs, \nu}}.
\end{equation}

The mass excess metric represents the relative amount of dust mass that is unaccounted for due to not only planetesimal formation but also as a result of the optically thick emission that arises from streaming instability-induced overdensities. Dust clumps of high surface density and/or large effective opacities can quickly enter the optically thick regime ($\tau_\nu^{\rm eff} \gg 1$) and could in principle hide a significant amount of mass when only (sub-)mm thermal emission is used to calculate the disk mass. Likewise, even a small number of planetesimals could skew the estimated mass lower as these planetesimals accrete while remaining optically inert in (sub-)mm wavelengths. We note that the gravitational collapse of planetesimals reduces the true dust mass over time due to subsequent pebble accretion, thus we also include the mass of the sink particles (i.e., planetesimals) in $m_{\rm true}$ when computing the mass excess.

A list of variables and parameters adopted in this work is presented in Table~\ref{tab:parameter_table}. In the following section we present the results from the radiative transfer and the resulting mass excess. By quantifying both the output intensity and the optical depth across the disk, we analyze the validity of the optically thin approximation (Equation~\ref{eq:sigma_d_eqn}) for estimating disk dust masses around Class II YSOs from (sub-)mm dust emission alone.

\section{Results}
\label{sec:results}

\begin{figure*}[ht!]
    \centering
    \includegraphics[width=\textwidth]{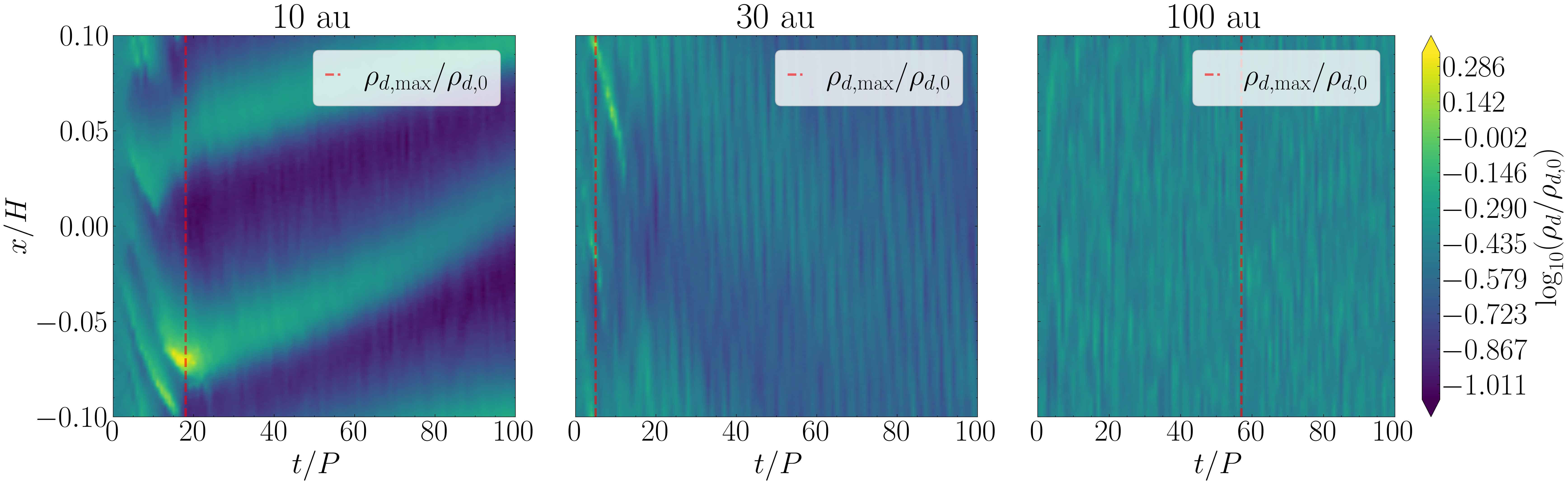}
    \caption{Time evolution of the dust distribution. The panels show the vertically and azimuthally averaged dust density as a function of time for each simulation and radial location. The red dashed line indicates the orbital period of maximum dust density.
    }
    \label{fig:densities_1}
\end{figure*}

\begin{figure*}[ht!]
    \centering
    \includegraphics[width=\textwidth]{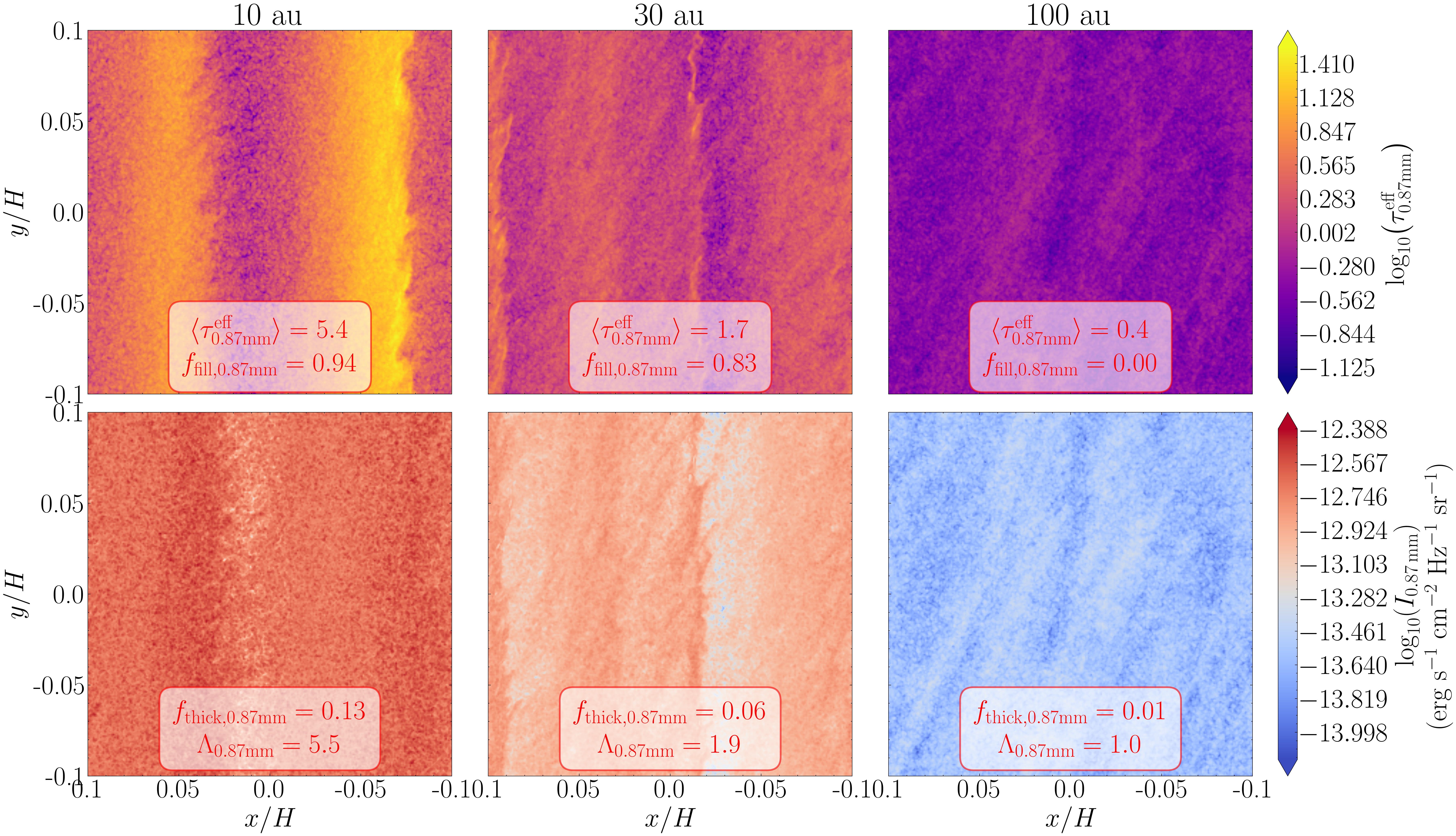}
    \caption{Radiative transfer results at the 0.87~mm wavelength. Each column corresponds to a different simulation and radial location in the disk. The top and bottom rows show the optical depth and the emergent intensity at the output plane ($z = L_z/2$), respectively, at the time of maximum dust density (red dashed lines in Fig.~\ref{fig:densities_1}).}
    \label{fig:densities_2}
\end{figure*}

\begin{figure*}
    \centering
    \includegraphics[width=\textwidth]{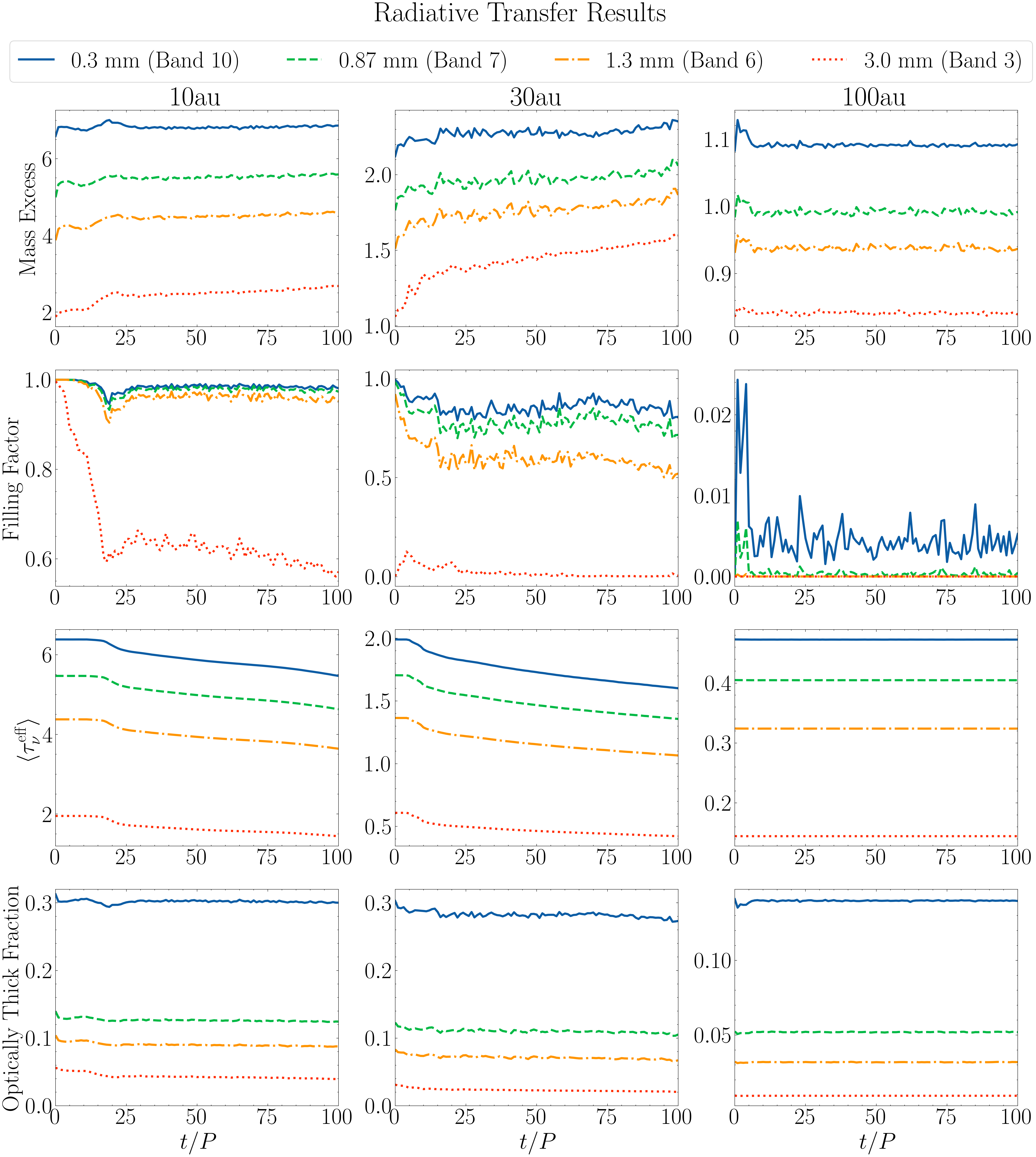}
    \caption{Observational diagnostics from our three streaming instability simulations, presented column-wise. The top plots display the mass excess over time and as function of both radial location and observed (sub-)mm wavelengths. The corresponding filling factors, average optical depths, and the optically thick fractions are shown in the lower rows, respectively.}
    \label{fig:sca}
\end{figure*}

In this section we present the results of our radiative transfer analysis, investigating how the streaming instability affects the outgoing intensity and, consequently, the disk dust masses as inferred from (sub-)mm observations. Since we focus on the impact of optically thick emission in localized dust overdensities, considering dust scattering is especially important as this can significantly reduce the specific (sub-)mm intensities in disk regions where $\tau_\nu^{\rm eff}\gg1$ \citep{Dullemond_2018, Zhu_2019, Sierra+20}. In such cases, scattering effects result in more emission per mass being required to match the observed flux. We therefore consider both absorption and scattering opacities in our radiative transfer calculations; with the analysis without scattering shown for comparison in the Appendix. 

Fig.~\ref{fig:densities_1} illustrates the temporal evolution of dust density at the three disk locations considered in our simulations. These projections depict the vertically and azimuthally averaged dust density, with red dashed lines indicating the orbital period of maximum density, which for the 10 and 30~au case roughly coincides with the maximum number of formed planetesimals. All three panels share a common color scale, shown in the rightmost panel. At 10 and 30~au, the simulations develop distinct overdense filaments that migrate inward over time, and that enhances the dust surface density by a factor of $\sim$2. While individual particles drift inward under gas drag, the apparent motion of the filaments reflects the collective dynamics of the dust and can differ in direction. This behavior is demonstrated in the space-time plot of the 10~au simulation, which reveals a distinct pattern speed. In this case, the filaments exhibit a slower inward drift than the particles contained within them, producing the illusion of an outward-moving pattern due to the decoupling between the pattern speed and the motion of individual particles.

The simulation at 100~au exhibits weak dust concentration, with the density remaining relatively uniform over time. This is consistent with the top-right panel of Fig.~\ref{fig:simulations}, which shows that even the largest grains remain at densities roughly 20 times below the 2$\rho_H$ threshold required for gravitational collapse. 

Fig.~\ref{fig:densities_2} displays Equation~\ref{eq:optical_depth} (top row) and Equation~\ref{eq:rad_transfer} (lower row) at the 0.87 mm wavelength, shown at the time of maximum density for each simulation. Each optical depth map corresponds to a vertical slice at $z=L_z/2$, with the mean ($\langle\tau_{0.87 \rm mm}^{\rm eff}\rangle$) and associated filling factor noted on the respective maps. Similarly, the outgoing intensity maps include the optically thick fraction at each location (Equation~\ref{eq:ff_thick}), alongside the corresponding mass excess (Equation~\ref{eq:me_equation}). The snapshots in  Fig.~\ref{fig:densities_2} reveal a physical separation of $\sim0.1H$ between the filaments formed by the streaming instability and illustrate the relationship between optical depth and the resulting thermal (sub-)mm emission. In the optically thick regime ($\langle\tau_{0.87 \rm mm}^{\rm eff}\rangle\geq1$), we find that the average optical depth is comparable to the corresponding mass excess (i.e., $\langle\tau_\nu^{\rm eff}\rangle\sim\Lambda_\nu$). 

The optical depth and corresponding intensities are calculated for all three simulations at every orbital period, with the resulting time series presented in Fig.~\ref{fig:sca}. The figure is organized column-wise, with the 10~au simulation shown in the left column, the 30~au case in the middle column, and the 100~au simulation in the right column. Each panel displays four lines, each corresponding to a different observational wavelength. The top row displays the mass excess, while the second row shows the filling factor. By the end of the 10~au simulation run, we find that mass excess is $\sim3-7$, increasing with the frequency of the wavelength. Similar to \citet{Scardoni+21}, we find that the streaming instability, if actively concentrating dust grains, always yields a mass excess. 

We observe the largest rate of increase in the mass excess before $t/T\sim25$ at 10 \& 30~au. This occurs during the early episode of strong clumping and planetesimal formation in the simulations. Following this initial burst, the long-term evolution of the mass excess is primarily driven by the gradual accumulation of mass into larger aggregates that remain untraceable in (sub-)mm observations. These results highlight how accretion by planetesimals increases the mass excess over time, and how its overall magnitude is driven primarily by scattering effects and the initial disk conditions.

At 10~au we observe an inverse correlation between the mass excess and the filling factor at all wavelengths. In this case, the (sub-)mm disk emission is optically thick, as shown in the first column, third row of Fig.~\ref{fig:sca}, where $\langle\tau_{\nu}^{\rm eff}\rangle>1$ at all frequencies. The strong clumping that occurs at $t/T\sim20$ (top left panel of Fig.~\ref{fig:simulations}) coincides with the sharpest decrease in the average optical depth and yields the lowest filling factors. This depicts a scenario in which the strong clumping by the streaming instability effectively reduces the dust column densities of the optically thick, adjacent environment, thereby reducing the filling factor. While planetesimal formation and the onset of pebble accretion reduce the optical depth over time, the emission remains optically thick throughout, even at the longest wavelength in our analysis (3~mm), where the mean optical depth is $\lesssim2$. The optically thick fractions shown in the lower panel illustrate how the streaming instability reduces the optically thick fraction, with a noticeable drop coinciding with the period of active planetesimal formation. However, we note that this trend is less pronounced than the other two diagnostics ($f_{\rm fill, \nu}$ \& $\langle\tau_{\nu}^{\rm eff}\rangle$). The 10~au case thus represents a disk environment in which the (sub-)mm emission remains optically thick, resulting in deviations from ideal blackbody emission of $\sim5-30$\%.

At 30~au (Fig.~\ref{fig:sca}, middle column), the mass excess is smaller than the 10~au case by a factor of $\sim2-3$ across all frequencies. In this scenario, we observe a similar correlation between mass excess and the filling factor at all wavelengths except the 3~mm (ALMA Band 3). At this longer wavelength, the filling factor increases with mass excess during the episode of planetesimal formation. We note that the inverse correlation emerges only when the mean optical depth is $\langle\tau_\nu^{\rm eff}\rangle\gtrsim1$, as in the optically thick regime the streaming instability has the effect of making the domain more optically thin. In contrast, the 3~mm emission remains optically thin throughout the simulation, with $\langle\tau_{3 \rm mm}^{\rm eff}\rangle\sim0.5$, resulting in the lowest filling factors of $f_{\rm fill, 3 \rm mm}\lesssim0.2$. Despite the low filling factors and $\langle\tau_{3 \rm mm}^{\rm eff}\rangle<1$, the disk mass is still underestimated, with $\Lambda_{3 \rm mm}\gtrsim1.5$. Moreover, although the mean optical depth at 30~au is approximately three times lower than at 10~au, the optically thick fractions at both locations are comparable. This example thus demonstrates that significant deviations between observed and true dust mass can occur even in the absence of uniformly optically thick conditions.

The simulation at 100~au (Fig.~\ref{fig:sca}, right column) is particularly interesting, as it produces no planetesimals, yet shows that the mass excess can fall below unity. This occurs at the 3~mm wavelength, where the optical depth, filling factor, and the optically thick fraction are lowest, with $\langle\tau_{3 \rm mm}^{\rm eff}\rangle<0.2$, $f_{\rm fill, 3 \rm mm}\sim0$, and $f_{\rm thick, 3 \rm mm}\sim0.01$. In this regime, the observed flux slightly overestimates the true dust mass, yielding $\Lambda_{3 \rm mm} \sim 0.85$. Although the simulation at 100~au does not produce large-scale filamentary structures, the streaming instability still gives rise to local regions of overdensity, which at $0.3-0.87$~mm wavelengths can still yield non-zero filling factors. 

In this case, mass excess values less than unity arises because, in optically thin environments with high single-scattering albedos ($\omega_\nu \gtrsim 0.7$), scattering contributes positively to the emergent intensity. In such cases, thermal photons scattered from deeper layers can escape more easily, effectively boosting the observed intensity beyond what would be expected from absorption alone. This effect is most pronounced at 1.3~mm and 3~mm, where the albedos are $\gtrsim 0.9$ (Fig.~\ref{fig:opacities_and_albedo_1}) and the emission remains fully optically thin ($f_{\rm fill,\nu}\sim0$). This result is consistent with the findings of \citet{Sierra+20}, who showed that dust scattering can enhance (sub-)mm intensity when albedos are high ($\omega_\nu \gtrsim 0.6$) and optical depths are low ($\tau_\nu \sim 10^{-1} - 10^{-2}$). Similarly, \citet{Zhu_2019} reported that in this regime, traditional mass estimates can slightly overestimate the true dust mass, consistent with our findings of $\Lambda_{1.3 \rm mm} \sim 0.95$ and $\Lambda_{3 \rm mm} \sim 0.85$. At higher frequencies where this effect is not observed, such as at 0.3~mm where $\langle \tau_{\rm 0.3mm} \rangle \lesssim 0.5$ and $f_{\rm fill,\rm 0.3mm} \sim 0$, the mass error remains relatively low, with $\Lambda_{0.3 \rm mm} \sim 1.1$. Therefore, in such optically thin environments, the mass excess is low and comparable to the absorption-only case, which we present in the Appendix and Fig.~\ref{fig:abs}.

\section{Discussion}
\label{sec:discussion}

In this study, we examine how dust clumping driven by the streaming instability can lead to significant underestimations of protoplanetary disk masses when (sub-)mm fluxes are interpreted under the assumption of optically thin emission. Using high-resolution simulations coupled with radiative transfer modeling that includes both absorption and scattering, we quantify the extent to which the emergent intensity deviates from idealized expectations. We find that even a small number of overdense filaments can obscure a substantial fraction of the disk mass.

Although planetesimal formation and accretion processes further reduce the amount of mass traceable in (sub-)mm surveys, our results indicate that the dominant factors driving the mass excess are the initial disk conditions and the role of scattering. This aligns with the findings of \citet{Liu_2022}, who reported that the mass excess is heavily dependent on the dust surface density of the disk. It is also consistent with earlier studies highlighting the significant impact of scattering on observed (sub-)mm flux densities \citep{Zhu_2019, Liu+19, Sierra+20}. While the merging of planetesimals in our simulations may not fully reflect realistic accretion rates, it demonstrates that planetesimal growth can concentrate mass into optically thick regions that are effectively invisible to (sub-)mm observations, further compounding the mass budget problem.

We find that overdense filaments formed via the streaming instability frequently reach optical depth $\tau_\nu^{\rm eff} > 1$ at (sub-)mm wavelengths. When emission from such regions is interpreted as optically thin, the estimated dust mass leads to mass excess of $\Lambda_\nu\sim2-7$, depending on the observing frequency as well as the local disk conditions such as the surface density and temperature. Scattering significantly contributes to this effect, particularly at higher frequencies (e.g., ALMA Band 10), where large single scattering albedos ($\omega_\nu\gtrsim0.9$) substantially reduce the emergent (sub-)mm intensity, leading to optically thick fractions as high as $f_{\rm thick, \nu}\sim0.3$.

Our results show that the streaming instability can either increase or decrease the local optical depth and the associated filling factor, depending on the background disk conditions. In initially optically thick environments, dust clumping into filaments removes material from the surrounding medium, reducing the overall column density and making the region more optically thin, which lowers the filling factor. Conversely, in optically thin environments, the formation of localized overdensities increases the filling factor as more area exceeds $\tau_\nu^{\rm eff}>1$. In both cases, the resulting mass excess values can be comparable, showing how considerable amounts of mass can be effectively hidden from (sub-)mm observations, even when only a small fraction of the disk becomes optically thick.

We find that even low filling factors ($f_{\rm fill, \nu} \lesssim 0.05$) can lead to mass underestimations comparable to those in disks with much higher filling factors ($f_{\rm fill, \nu} \sim 0.8$), highlighting that a small number of optically thick structures -- such as several dust filaments -- can significantly bias mass estimates. In some cases, particularly at larger disk radii (e.g., 100~au), the mass excess can fall below unity at longer (sub-)mm wavelengths (Fig.~\ref{fig:sca}, top-right panel). The net positive flux contribution in these cases results in mass overestimates of up to 15\%. This effect arises from the positive contribution of scattering that occurs in optically thin regimes with high albedos (Fig.~\ref{fig:opacities_and_albedo_1}), a behavior also noted in previous studies \citep{Zhu_2019, Sierra+20}. In such cases, scattering redirects thermal photons from deeper layers into the line of sight, slightly boosting the emergent intensity and leading to an overestimate of the dust mass. This demonstrates the important role of scattering even when the disk is optically thin, further calling into question the reliability of flux-based disk mass estimates calculated using Equation~\ref{eq:sigma_d_eqn}; which assumes a position-independent source function and thus considers only the absorption opacities.

Our results complement recent studies by \citet{Rucska_2023} and \citet{Scardoni+21}, who similarly found that strong clumping via the streaming instability can drive mass excess of order unity or higher. Additionally, comparisons between disk masses derived from the optically thin approximation and those inferred from full spectral energy distribution modeling have shown that flux-based dust mass estimates typically underestimate the true solid content by factors of $\sim1.5-6$ \citep{Ballering+19, Ribas+20, Macias+21, Xin+23, Rilinger_2023}. Our findings are consistent with these results and support the hypothesis that the ``missing'' mass in Class II disks could be a consequence of both early planetesimal formation and/or optically thick emission driven by localized dust concentrations \citep{Savvidou+24}.

The correction factors we report are consistent with the discrepancies reported in \citet{Manara_2018}, who found a mass budget discrepancy of $\sim3-5$ for systems around stars with masses $<2M_\odot$. A similar population study by \citet{Mulders+21} found that, when correcting for observational biases, this discrepancy largely disappears, with disk and exoplanet masses around solar mass stars ($0.5<M_*/M_\odot<2$) being of similar magnitude. For this to resolve the mass budget problem, however, a planet formation efficiency close to unity would be required. If planet formation is inefficient, this would imply mass excess factors exceeding unity, with efficiencies of $\sim$30–50\% leading to mass underestimates comparable to those reported in our work. Our results therefore align with a growing consensus that a combination of optically thick emission, dust growth beyond mm sizes, and early planetesimal formation is needed to reconcile disk observations with mature planetary systems.

As the choice of dust model directly determines the dust grain opacities and (sub-)mm emissivities, assumptions about grain composition can significantly affect the inferred disk mass and, consequently, the mass excess. In this work, we consistently adopt the DSHARP dust model, ensuring compatibility with recent ALMA-based analyses. However, alternative models such as the DiscAnalysis project (DIANA; \citealt{Woitke+16}) are worth consideration in the context of the streaming instability. Disks modeled with DIANA opacities tend to be more optically thick than those using DSHARP, leading to systematically larger mass excess values. For example, \citet{Liu_2022} showed that, under the optically thin assumption, switching from DSHARP to DIANA increases the inferred mass excess by a factor of $\sim$2. Additional factors such as grain porosity and disk inclination can further amplify this effect, with more porous grains or more inclined viewing angles leading to larger optical depths and mass excess values \citep{Liu_2022, Liu+24}. Future work should incorporate a broader set of dust opacity models and perform radiative transfer calculations over a range of inclinations to better constrain the combined impact of dust properties and viewing geometry.

\section{Conclusions}
\label{sec:conclusion}

We have used high-resolution simulations of the streaming instability combined with radiative transfer modeling to quantify how dust clumping and planetesimal formation affect the (sub-)mm flux and inferred dust masses of protoplanetary disks. Our findings further highlight the limitations of the optically thin approximation commonly used in observational analyses, emphasizing the role of scattering by (sub-)mm dust grains on the emergent intensity. The main conclusions of our work are as follows:

\begin{enumerate}
    \item Dust overdensities from the streaming instability can become optically thick at (sub-)mm ALMA wavelengths, leading to mass excess of $\Lambda_\nu \sim 2 - 7$ in the inner disk regions ($\lesssim$30~au).
    \item Dust scattering increases the effective optical depth and suppresses emergent intensities at (sub-)mm wavelengths, yielding larger mass excess. Including scattering raises $\Lambda_\nu$ by factors of $\sim1.5 - 3$ compared to absorption-only calculations shown in the Appendix. 
    \item The redistribution of grains due to streaming instability-induced clumping increases the filling factors in initially optically thin environments and decreases it in optically thick regions.
    \item Significant dust mass underestimations can occur even if only a small fraction of the disk area is optically thick, demonstrating how relatively few structures can effectively hide a lot of disk mass when (sub-)mm flux densities are used to estimate the dust mass.
    \item Even when the mean optical depth is less than unity and the filling factors are low, the presence of overdensities can make the optically thin approximation unreliable when the single-scattering albedos are high.
\end{enumerate}

Our results suggest that the missing mass problem in protoplanetary disks can be partially attributed to localized, optically thick emission caused by dust clumping and early planetesimal formation. Observational studies that rely on integrated (sub-)mm fluxes and assume optically thin conditions are thus underestimating the true disk masses, particularly in regions where the streaming instability is active. 

Dust scattering also contributes significantly to this reduction in (sub-)mm emission, except in optically thin cases where scattering by large $\sim$~cm-sized grains actually yields a net positive contribution. In such instances, the flux per mass can be overestimated by up to $\sim$15\%. Otherwise, we find that optically thick emission yields erroneous mass underestimates if the disk is assumed to be optically thin throughout, with mass excess of $\sim2-7$ at 10~au and $\sim1.5-2.4$ at 30~au (Fig.~\ref{fig:sca}). Even without the effects of scattering, overdensities result in mass excess of $\sim2-3$ at 10~au and $\sim1.4-1.6$ at 30~au (see Appendix). 

Long-wavelength facilities such as the next generation Very Large Array (ngVLA; \citealt{Murphy+18}) will be critical in probing optically thin emission at high resolution and will help reveal dust populations currently missed by ALMA (see e.g., \citealt{Ricci+18}). Future work should incorporate models that account for dust dynamics such as (sub-)mm dust growth, early planetesimal formation, and self-consistent three-dimensional radiative transfer. To support this, we release the accompanying codebase, \texttt{protoRT} \citep{protoRT}, which provides a flexible, open-source framework for modeling radiative transfer and assessing dust mass underestimation in protoplanetary disk simulations. The release includes all code and data necessary to reproduce our analysis, along with comprehensive documentation. Our results highlight how physical processes such as the streaming instability directly impacts the observed thermal emission, and in turn the amount of dust mass that is unaccounted for under idealized assumptions.

\section*{Acknowledgements}

We thank Leonardo Krapp for insightful discussions. W.L, C-C. Y., and J.B.S acknowledge support from NASA under the Theoretical and Computational Astrophysical Networks (TCAN) grant \# 80NSSC21K0497. W.L. also acknowledges support from NASA under the Emerging Worlds (EW) program grant \#80NSSC22K1419 and NSF via grant AST-2007422. J.B.S also acknowledge support from NASA under Exoplanets Research Program (XRP) grant \# 80NSSC22K0267. C.C.Y.\ is also grateful for the support from NASA via the Emerging Worlds program (grant \#80NSSC23K0653) and the Astrophysics Theory Program (grant \#80NSSC24K0133). J.L. acknowledges support from NASA under the Future Investigators in NASA Earth and Space Science and Technology grant \# 80NSSC22K1322. 

The simulations were performed on Stampede at NSF's Texas Advanced Computing Center (TACC) using XSEDE/ACCESS grant TG-AST140014, and the Discovery cluster at New Mexico State University \citep{TrecakovVonWolff21}. This work utilized resources from the New Mexico State University High Performance Computing Group, which is directly supported by the National Science Foundation (OAC-2019000), the Student Technology Advisory Committee, and New Mexico State University and benefits from inclusion in various grants (DoD ARO-W911NF1810454; NSF EPSCoR OIA-1757207; Partnership for the Advancement of Cancer Research, supported in part by NCI grants U54 CA132383 NMSU). This research was made possible by the open-source projects \texttt{jupyter} \citep{Jupyter}, \texttt{IPython}
\citep{IPython}, \texttt{matplotlib} \citep{matplotlib1, matplotlib2}, \texttt{NumPy} \citep{numpy}, \texttt{SymPy} \citep{sympy}, and \texttt{AstroPy} \citep{Astropy_2022}.
 
\appendix 

\section{Absorption Only Scenario}
\label{sec:appendix}

\begin{figure*}
    \centering
    \includegraphics[width=\textwidth]{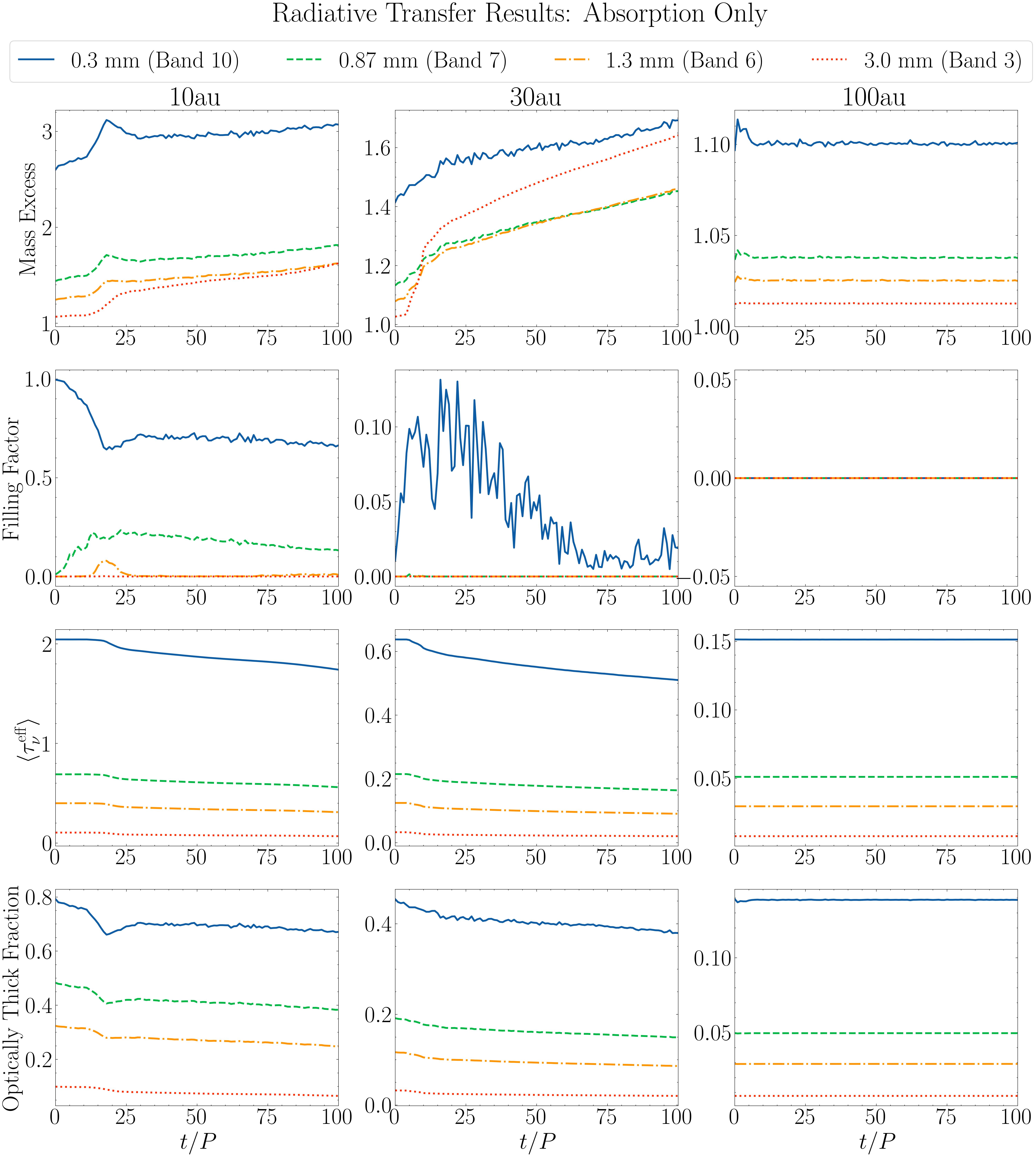}
    \caption{Absorption-only case. The results from the radiative transfer calculations from our three streaming instability simulations are presented column-wise. The top plots display the mass excess over time and as function of both radial location and observed (sub-)mm wavelengths. The corresponding filling factors, average optical depths, and the optically thick fractions are shown in the lower rows, respectively.}
    \label{fig:abs}
\end{figure*}

To isolate the impact of scattering, we perform our radiative transfer analysis excluding the scattering opacities. This comparison allows us to assess how much scattering alone modifies the inferred disk mass and optical depth structure. In this absorption-only case, the effective opacity reduces to $\kappa_\nu^{\rm eff} = \kappa_\nu$, and the source function becomes spatially constant ($S_\nu = B_\nu$), removing the redistribution effects introduced by scattering.

As expected, the simulation at 100~au still remains optically thin throughout when scattering is excluded. As shown in the top-right panel Fig.~\ref{fig:abs}, the mass excess stays close to unity across all wavelengths, indicating negligible mass bias. The corresponding filling factors shown in the panel below are near zero for all bands, and the mean optical depth shown below that are low ($0.007 \lesssim \langle \tau^{\rm eff}_{\nu} \rangle \lesssim 0.15$). At 0.3~mm, where the optical depth is the largest, the resulting mass excess is $\Lambda_{0.3 \rm mm}\sim$1.1. The optically thick fractions shown in the lower-right panel remain relatively constant at every orbit and are in agreement with optically thin conditions ($f_{\rm thick,\nu}\sim\langle \tau^{\rm eff}_{\nu} \rangle$). This scenario serves as a baseline for understanding the pure optically thin limit and demonstrates how, in the absence of (sub-)mm scattering by large grains, the mass excess never drops below unity and optically thick fractions scale with the optical depth.

At 10~au, where strong dust clumping and planetesimal formation occur, the mass excess under the absorption-only assumption reaches up to $\Lambda_{0.3 \rm mm} \sim 3$ during peak overdensity (Fig.~\ref{fig:abs}, top left). While still significant, this is lower than in the full radiative transfer model that includes scattering by a factor of $\sim$2. Although resulting optical depths are lower by comparable factors, the filling factors are significantly lower, being less than 0.2 at all bands except the 0.3~mm which by the end of the simulation is $f_{\rm fill,0.3mm}\lesssim0.7$. Even though photons emitted from optically thick clumps are still partially absorbed, without scattering redirecting photons out of the line of sight, the reduction in the emergent intensity is lower. As a result, the inferred dust mass is closer to the true value. The simulation at 30~au (Fig.~\ref{fig:abs}, top-middle panel) represents a transitional regime. Without scattering, we observe a moderate mass excess ($\Lambda_{0.3 \rm mm} \sim 2$) and optically thick emission primarily at the shorter wavelength. At longer wavelengths, the disk remains mostly optically thin with $f_{\rm fill,\nu}\sim0$, but still shows non-negligible mass excess due to localized clumping by the streaming instability. Despite the low filling factors, the mass excess remains above 1.4, showing how dust overdensities can bias the inferred mass estimates even in optically thin conditions.

At 10~au and 0.3~mm, the disk initially appears uniformly optically thick, with a filling factor near unity. As the streaming instability concentrates material into narrow filaments, the average optical depth decreases and the filling factor drops to $\sim$0.7. This demonstrates how clumping reduces the column density in between the filaments and can in certain cases reduce the filling factor. At longer wavelengths such as 0.87~mm, where the disk is initially optically thin, the opposite trend occurs, with clumping increasing the local column density within filaments, raising the filling factor. 

This comparison emphasizes how scattering can substantially change the optical depth of the disk and in turn the emergent thermal emission. Even modest optical depths of $0.5 - 1$, when combined with high albedos ($\omega \gtrsim 0.9$), can suppress the emergent flux. This behavior was explained by \citet{Miyake_1993}, whose isothermal, plane-parallel solution with isotropic scattering shows that the emergent intensity decreases with increasing albedo at same optical depth. This effect is attributed both to the reduction of true emissivity at same optical depth as albedo increases, and also to higher prevalence of scattering photons out of the line of sight than into the line of sight, reducing $I_\nu$ and artificially lowering the filling factor and optically thick fractions. Consequently, when scattering is included, inferred disk masses are significantly lower than in the absorption-only scenario. At 10~au and 0.3 mm, for instance, $\Lambda_\nu$ rises from $\sim$3 to $\sim$7 purely due to the inclusion of scattering. This increases the mass excess by factors of $\sim$1.5–3, underscoring the need to account for scattering when interpreting (sub-)mm fluxes from disks.

The simulations at 10 and 30~au further reveal that the temporal evolution of the mass excess in this context is most sensitive to planetesimal formation, as we observe significantly steeper increases during the planetesimal formation stage when compared to the analysis in Fig.~\ref{fig:sca} that includes scattering. This suggests that in the absence of scattering, the formation and growth of planetesimals play a more dominant role in increasing the mass excess. When scattering is included, the increase in the mass excess during the same period is more modest, indicating that scattering suppresses the emergent flux more uniformly across time. This distinction highlights how scattering can mask the signatures of planetesimal formation and dominate the overall contribution to the mass excess. As a result, neglecting scattering can lead to overestimating the role of dynamical processes alone in producing the observed mass excess.

\bibliographystyle{aasjournalv7}

\bibliography{Bibliography} 
\end{CJK}
\end{document}